\documentclass[draft,a4paper,onecolumn]{IEEEtran}
\usepackage{amsfonts, amssymb,amsmath}
\usepackage{amscd}
\usepackage{color}
\usepackage{iftex}
\ifPDFTeX
   \usepackage[utf8]{inputenc}
   \usepackage[T1]{fontenc}
   \usepackage{lmodern}
\else
   \ifXeTeX
     \usepackage{xltxtra}
   \else
     \usepackage{luatextra}
   \fi
   \defaultfontfeatures{Ligatures=TeX}
\fi

\newtheorem{theo}{Theorem}[section]

\newtheorem{coro}[theo]{Corollary}
\newtheorem{prop}[theo]{Proposition}
\newtheorem{lemm}[theo]{Lemma}

\newtheorem{ex}[theo]{Example}

\newcommand{\Z}{{\mathbb{Z}}}
\newcommand{\F}{{\mathbb{F}}}

\newcommand{\vv}{{\bf  v}}
\newcommand{\vb}{{\bf  b}}
\newcommand{\vw}{{\bf  w}}

\newcommand{\vx}{{\bf  x}}
\newcommand{\vh}{{\bf  h}}
\newcommand{\ve}{{\bf  e}}
\newcommand{\vk}{{\bf  k}}

\newcommand{\K}{{\cal K}}
\newcommand{\zero}{{\mathbf{0}}}
\newcommand{\one}{{\mathbf{1}}}
\newcommand{\two}{{\mathbf{2}}}

\newcommand{\gh}{\operatorname{GH}}
\newcommand{\rank}{\operatorname{rank}}

\newcommand{\bfomega}{\boldsymbol{\omega}}

\newcommand{\GF}{\operatorname{GF}}
\definecolor{gold}{rgb}{0.85,.26,0}

\title{Ranks and Kernels of Codes from Generalized Hadamard Matrices\thanks{This work was partially supported by the
Spanish MICINN under Grant TIN2013-40524-P, and
by the Catalan AGAUR under Grant 2014SGR-691.\newline \hspace*{0.3cm} $^1$Steven T. Dougherty is with the Department of Mathematics, University of Scranton, Scranton PA 18510, USA. \newline \hspace*{0.3cm} $^2$Josep Rif\`{a} and Merc\`{e} Villanueva
are with the Department of Information and Communications
Engineering, Universitat Aut\`{o}noma de Barcelona, 08193 Cerdanyola del Vall\`{e}s, Spain.}}

\author{Steven T. Dougherty$^1$, Josep Rif\`{a}$^2$ and Merc\`{e} Villanueva$^2$}

\date{\today}

\begin{document}

\maketitle

\begin{abstract}
The ranks and kernels of generalized Hadamard matrices are studied.
It is proven that any generalized Hadamard matrix $H(q,\lambda)$ over $\F_q$, $q>3$, or $q=3$ and $\gcd(3,\lambda)\not =1$, generates a
self-orthogonal code. This result puts a natural upper bound on the rank of the generalized Hadamard matrices.
Lower and upper bounds are given for the dimension of the kernel  of the corresponding generalized Hadamard codes.
For specific ranks and dimensions of the kernel within these bounds, generalized Hadamard codes are constructed.
\end{abstract}

\section{Introduction}

Let $\F_q=\GF(q)$ denote the finite field with $q$ elements, where $q=p^e$, $p$ prime.
Let $\F^n_q$ be the vector space of dimension $n$ over $\F_q$.
The {\it Hamming distance} between vectors $\vw$, $\vv \in \F^n_q$,
denoted by $d(\vw,\vv)$, is the number of coordinates in which $\vw$ and
$\vv$ differ. 
A {\it code} $C$ over $\F_q$ of length $n$ is a nonempty subset of $\F^n_q$. The
elements of $C$ are called {\it codewords}. A code $C$ over $\F_q$ is called {\it
linear} if it is a linear space over $\F_q$ and, it is called {\it $K$-additive}
if it is a linear space over a subfield $K \subset \F_q$.
The dimension of a $K$-additive code $C$ over $\F_q$ is defined as the number $k$ such that
$q^k=|C|$. Note that $k$ is not necessarily an integer, but $ke$ is an integer, where $q=|K|^e$.
The {\it minimum distance} of a code is the
smallest Hamming distance between any pair of distinct codewords.
Two codes $C_1$, $C_2 \subset \F^n_q$ are said to be {\it permutation equivalent}
if there exists a permutation $\sigma$ of the $n$ coordinates such that
$C_2=\{\sigma(c_1,c_2,\ldots,c_n)=(c_{\sigma^{-1}(1)},\ldots, c_{\sigma^{-1}(n)}) : (c_1,c_2,\ldots,c_n) \in C_1 \}$, \cite{CHLL97}, \cite{McWill}. Without loss of generality, we
shall assume, unless stated otherwise, that the all-zero vector, denoted by $\zero$, is
in $C$.

%

Two structural parameters of (nonlinear) codes are the dimension of
the linear span and the kernel. The {\it linear span} of a code $C$ over $\F_q$, denoted by ${\cal R}(C)$, is the
subspace over $\F_q$ spanned by $C$, that is ${\cal R}(C)=\langle C \rangle$.
The dimension of ${\cal R}(C)$ is called the {\it rank} of $C$ and is denoted by $\rank(C)$.
If $q=p^e$, $p$ prime, we can also define ${\cal R}_p(C)$ and $rank_p(C)$ as the subspace
over $\F_p$ spanned by $C$ and its dimension, respectively.
The {\it kernel} of a code $C$ over $\F_q$, denoted
by ${\cal K}(C)$, is defined as ${\cal K}(C) =\{\vx\in \F_q^n \ : \ \alpha \vx+C=C$ for all $\alpha \in \F_q  \}.$
If $q=p^e$, $p$ prime, we can also define the {\it $p$-kernel} of $C$ as
$\K_p(C)=\{ \vx\in \F_q^n \ : \ \vx+C=C \}.$  Since we assume that $\zero \in C$, then
$\K(C)$ is a linear subcode of $C$ and $\K_p(C)$ is an $\F_p$-additive subcode.
We denote the dimension of the kernel and
$p$-kernel of $C$ by $\ker(C)$ and $\ker_p(C)$, respectively.
These concepts were first defined
in \cite{merce1} for codes over $\F_q$, generalizing the binary case described previously in \cite{BGH}, \cite{PhLV}.
In \cite{merce1}, it was proved that the code $C$ over $\F_q$ can be written as the union of
cosets of $\K(C)$ (resp. $\K_p(C)$), and $\K(C)$ (resp. $\K_p(C)$) is the largest such linear code
over $\F_q$ (resp. $\F_p$) for which this is true. Moreover, it is clear that $\K(C) \subseteq \K_p(C)$.

\medskip
A {\it generalized Hadamard} ($\gh$) {\it matrix} $H(q,\lambda)=(h_{ij})$ of order $n=q\lambda$
over $\F_q$ is a $q\lambda \times q\lambda$ matrix with entries from
$\F_q$ with the property that for every $i,j$, $1\leq i<j\leq q\lambda$, each of the multisets
$\{ h_{is} - h_{js} :  1 \leq s \leq q\lambda \}$ contains every element of $\F_q$ exactly $\lambda$ times.
It is known that since $(\F_q,+)$ is an abelian group then $H(q,\lambda)^T$
is also a $\gh$ matrix, where $H(q,\lambda)^T$ denotes the transpose of $H(q,\lambda)$ \cite{jun}.
An ordinary Hadamard matrix of order $4\mu$ corresponds to a $\gh$ matrix $H(2,\lambda)$ over $\F_2$, where $\lambda=2\mu$.

Two $\gh$ matrices $H_1$ and $H_2$ of order $n$ are said to be {\it equivalent}
if one can be obtained from the other by a permutation of the rows and columns
and adding the same element of $\F_q$ to all the coordinates in a row or in a column.
We can always change the first row and column of a $\gh$ matrix
into zeros and we obtain an equivalent $\gh$
matrix which is called {\it normalized}. From a normalized Hadamard matrix $H$,
we denote by $F_H$ the code over $\F_q$ consisting of the rows of $H$,
and $C_H$ the one defined as $C_H=\bigcup_{\alpha \in \F_q} (F_H+\alpha {\bf 1})$, where $F_H+\alpha {\bf 1}=\{ \vh+ \alpha {\bf 1} : \vh \in F_H\}$ and ${\bf 1}$ denotes the all-one vector.
The code $C_H$ over $\F_q$ is called {\it generalized Hadamard} ($\gh$) {\it code}. Note that $F_H$ and $C_H$ are generally nonlinear codes over $\F_q$.

To check whether two $\gh$ matrices are equivalent
is known to be an NP-hard problem. However, we can use the invariants related to the linear span and kernel
of the corresponding $\gh$ codes in order to help in their classification,
since if two GH codes have different ranks or dimensions of the kernel, the GH
matrices are nonequivalent.

\begin{lemm}\label{lemm:1}
Let $H$ be a $\gh$ matrix over $\F_q$. Then $\rank(C_H)= \rank(F_H) +1$ and $\ker(C_H) = \ker(F_H) +1$.
\end{lemm}
\begin{IEEEproof}
It is straightforward from the definitions.
\end{IEEEproof}

In this paper, we shall only study $\gh$ codes over a finite field because of the following proposition.

\begin{prop}
Let $H$ be a $\gh$ matrix over a ring $R$ that is not a field.  Then  $\K(F_H)$ is trivial.
\end{prop}
\begin{IEEEproof}
If $\vv $ is a row of $H$ and $\vv \in \K(F_H)$, then $\alpha \vv $ must be in $F_H$ for all $\alpha$ in the ring. If $\alpha$ is a non-unit, $\alpha \vv$ cannot have, as coordinates, every element of $R$ $\lambda$ times and so $\alpha \vv$ is not in $ F_H. $  Hence the kernel is trivial.
\end{IEEEproof}

\medskip
The rank and dimension of the kernel for ordinary Hadamard codes over $\F_2$ have been studied in
\cite{HadPower2}, \cite{HadAnyPower}, \cite{HadAdditius}. Specifically, lower and upper bounds for these
two parameters were established, and the construction of an Hadamard code for all allowable ranks and
 dimensions of the kernel between these bounds was also given.

In this paper, we present a generalization of the above results.
The paper is organized as follows. In Section~\ref{sec:Kronecker}, we
give the implications for the rank and kernel for the standard Kronecker sum construction.
In Section~\ref{sec:kernel}, we find lower and upper bounds for the dimension of the kernel of a $\gh$ code,
constructing examples for some specific values in this interval.
In Section~\ref{sec:rank}, we establish an upper bound for the rank, by proving that the $\gh$ codes are
self-orthogonal unless $q=3$ and $\gcd(\lambda,q)=1$. Finally, in Section~\ref{sec:conclusion},
we give some conclusions and discuss further avenues of research on this topic.

\section{Kronecker sum construction}
\label{sec:Kronecker}

A standard method to construct $\gh$ matrices from other $\gh$ matrices
is given by the {\it Kronecker sum construction} \cite{seo}, \cite{shr}. That is,
if $H(q,\lambda)=(h_{ij})$ is any $q\lambda \times q\lambda$ $\gh$ matrix over $\F_q$, and $B_1, B_2,
\ldots, B_{q\lambda}$ are any $q\mu\times q\mu$ $\gh$ matrices over $\F_q$, then the
matrix in Table~\ref{kron} gives a $q^2\lambda \mu\times q^2\lambda \mu$ $\gh$ matrix over $\F_q$, denoted by $H \oplus [B_1,
B_2, \ldots, B_n]$, where $n=q\lambda$. If $B_1=B_2=\cdots =B_n=B$, then we write $H \oplus [B_1, B_2,
\ldots, B_n]=H \oplus  B$.

\begin{table}[ht] \caption{\label{kron} Kronecker sum construction}
 \centering $ H \oplus [B_1,
B_2, \ldots, B_n]=\left(%
\begin{array}{cccc}
  h_{11}+B_1 & h_{12}+B_1 & \cdots & h_{1n}+B_1 \\
  h_{21}+B_2 & h_{22}+B_2 & \cdots & h_{2n}+B_2 \\
  \vdots & \vdots & \vdots & \vdots \\
  h_{n1}+B_n & h_{n2}+B_n & \cdots & h_{nn}+B_n \\
\end{array}%
\right)$
\end{table}

Let $S_q$ be the normalized $\gh$ matrix $H(q,1)$ given by the multiplicative
table of $\F_q$. As for ordinary Hadamard matrices over $\F_2$, starting from a $\gh$ matrix $S^1=S_q$, we can
recursively define $S^t$ as a $\gh$ matrix $H(q,q^{t-1})$, constructed as $S^t=S_q \oplus
[S^{t-1},S^{t-1},\ldots,S^{t-1}]=S_q \oplus S^{t-1}$ for $t > 1$,
which is called a {\it Sylvester $\gh$ matrix}.

Let $\vv=(v_1,v_2,\ldots, v_n)$ and $\vw=(w_1,w_2,\ldots, w_m)$ be two vectors over $\F_q$.
Then, $\vv \oplus \vw=(v_1+w_1, \ldots, v_1+w_m, v_2+w_1, \ldots, v_2+w_m, \ldots, v_n+w_1,\ldots, v_n+w_m )$.

\begin{lemm}\label{lem:Kro1}
Let $H_1$ and $H_2$ be two $\gh$ matrices over $\F_q$ and $H=H_1 \oplus H_2$. Then
$\rank(C_H)=\rank(C_{H_1})+\rank(C_{H_2})-1$ and $\ker(C_H)=\ker(C_{H_1})+\ker(C_{H_2})-1$.
\end{lemm}

\begin{IEEEproof}
Without loss of generality, assume that $H_1(q,\lambda)$ and $H_2(q,\mu)$ are normalized $\gh$ matrices with row vectors $\vv_i \in {H_1}$ and $\vw_j \in {H_2}$, $i\in \{1,\ldots, q\lambda\}$ and $j\in \{1,\ldots, q\mu \}$. Since $\vv_i \oplus \vw_j= \vv_i\oplus \zero + \zero \oplus \vw_j$, it is easy to see that all row vectors in $H=H_1 \oplus H_2$ are linearly generated by $\vv_i\oplus \zero$ and $\zero \oplus \vw_j$. Moreover, since the two vectors $\vv_i \oplus \zero$ and
$\zero \oplus \vw_j $ are linearly independent for any $\vv_i\not =\zero$ and $\vw_j\not =\zero$, we have that $\rank(F_H) = \rank(F_{H_1}) + \rank(F_{H_2})$. By Lemma~\ref{lemm:1}, $\rank(C_H) = \rank(F_H)+1 = \rank(F_{H_1})+\rank(F_{H_2})+1= \rank(C_{H_1})+\rank(C_{H_2})-1$.

For the kernel, we have that $\vv_i\oplus \vw_j \in {\cal K}(F_H)$ if and only if $\vv_i\in {\cal K}(F_{H_1})$ and $\vw_j\in {\cal K}(F_{H_2})$. Hence, the vectors $\vv_i \oplus \zero$ and $\zero \oplus \vw_j$,  where $\vv_i\in {\cal K}(F_{H_1})$ and $\vw_j\in {\cal K}(F_{H_2})$, linearly generate ${\cal K}(F_H)$ and so, $\ker(F_H) = \ker(F_{H_1}) + \ker(F_{H_2})$. Finally, the result follows by Lemma~\ref{lemm:1}.
\end{IEEEproof}

\begin{coro} \label{coro:2.2}Let $B$ be a $\gh$ matrix over $\F_q$ and $H=S_q \oplus B$.
Then  $\rank(C_H)=\rank(C_B)+1$ and $\ker(C_H)=\ker(C_B)+1$.
\end{coro}

\begin{IEEEproof}
It follows directly from Lemma \ref{lem:Kro1} and the fact that
$C_{S_q}$ is a linear code of dimension 2, which means that $\rank(C_{S_q})=\ker(C_{S_q})=2$.
\end{IEEEproof}

\begin{prop}\label{lem:Kro1b}
Let $H_1$ and $H_2$ be two $\gh$ matrices over $\F_q$ with $H=H_1 \oplus H_2$. Let $K$ be a subfield of $\F_q$.
If $C_{H_1}$ and $C_{H_2}$ are $K$-additive, then $C_{H}$ is also $K$-additive and $\dim_K(C_H)=\dim_K(C_{H_1})+\dim_K(C_{H_2})-1$.
\end{prop}

\begin{IEEEproof}
It is straightforward using the same argument as in the proof of Lemma \ref{lem:Kro1}.
\end{IEEEproof}

\begin{lemm}\label{lem:KroSumRank} Let $H=H_1 \oplus [B_1,B_2, \ldots, B_n]$, where $H_1(q,\lambda)$ is a $\gh$ matrix over $\F_q$ and $B_1, B_2, \ldots, B_{n}$ are $\gh$ matrices over $\F_q$, where $n=q\lambda$. Then $\rank(C_H)=\rank(C_{H_1})+\dim(\langle C_{B_1} \cup C_{B_2} \cup \cdots \cup C_{B_n}  \rangle) -1$.
\end{lemm}
\begin{IEEEproof}
The row vectors in a GH matrix are linearly generated by some of its rows. Let $\vv_i$ be the $i$th generator row of $H_1$. Using the same argument as in the proof of Lemma~\ref{lem:Kro1}, we see that the vectors $\vv_i \oplus \zero$ and $\zero\oplus \vw_{jk}$ linearly generate $H$, where $\vw_{jk}$ is the $j$th generator row of $B_k$. Hence, we can conclude that $\rank(F_H)=\rank(F_{H_1})+\dim(\langle F_{B_1} \cup F_{B_2} \cup \cdots \cup F_{B_n}\rangle)$. Finally, by Lemma~\ref{lemm:1}, $\rank(C_H)=1+\rank(F_H)=1+\rank(F_{H_1})+\dim(\langle F_{B_1} \cup F_{B_2} \cup \cdots \cup F_{B_n}  \rangle)= \rank(C_{H_1})+ \dim(\langle C_{B_1} \cup C_{B_2} \cup \cdots \cup C_{B_n}  \rangle)-1$.
\end{IEEEproof}

\begin{coro}\label{coro:KroSumRank} Let $H=S_q \oplus [B_1,B_2, \ldots, B_q]$, where $B_1, B_2, \ldots, B_{q}$ are $\gh$ matrices over $\F_q$. Then $\rank(C_H)=\dim(\langle C_{B_1} \cup C_{B_2} \cup \cdots \cup C_{B_q}  \rangle) +1$.
\end{coro}
\begin{IEEEproof}
It is straightforward from Lemma \ref{lem:KroSumRank} and  the fact that $C_{S_q}$ is a linear
code of dimension 2.
\end{IEEEproof}

\begin{lemm}\label{lem:KroSumKernel} Let $H=S_q \oplus [B_1,B_2, \ldots, B_q]$, where $B_1, B_2, \ldots, B_{q}$ are $\gh$ matrices over $\F_q$. If there exists, at least, one matrix $B_i$ which is not translation equivalent to any other $B_j$ for $i, j\in\{1,2,\ldots,q\}$ and $i\not= j$, then $\K(C_H)=\{ \zero \oplus \vx=(\vx,\vx,\ldots,\vx) : \vx \in \K(C_{B_1}) \cap \K(C_{B_2}) \cap \cdots \cap \K(C_{B_q})\}$.
\end{lemm}

\begin{IEEEproof}
Note that the row vectors in $S_q$ are linearly generated by one of its rows, for instance $\ve$. Hence, the rows in $S_q$ can be described by $\ve_i= \omega^i \ve$, where $\omega$ is a primitive element in $\F_q$ and $i\in \{1,2,\ldots,q-1\}$, and $\ve_q=\zero$. Let $\vb_{jk}$ be the $j$th row of $B_k$. If $\ve_k\oplus \vb_{jk} \in \K(F_H)$, then adding $\ve_i\oplus B_i$ we obtain elements belonging to $C_H$
for all $i\in\{1,2,\ldots,q\}$. Hence, $\vb_{jk} + B_i = B_s$ for $i\in\{1,2,\ldots,q\}$, where $\ve_s=\ve_k+\ve_i$. Therefore, assuming the hypothesis in the statement, that is, $B_i$ is not translation equivalent to any other $B_s$ for $s\not =i$, we can conclude that when $\ve_k\oplus \vb_{jk} \in \K(F_H)$ we have that $\ve_k=\zero$ and
$\vb_{jq} \in \K(F_{B_1}) \cap \K(F_{B_2}) \cap \cdots \cap \K(F_{B_q})$. 

Finally, as the all-one vector $\one$ belongs to all the kernels $\K(C_{B_i})$, for $i\in\{1,2,\ldots,q\}$, and also to $\K(C_H)$, we obtain the statement.
\end{IEEEproof}

\section{Kernel dimension of GH codes}
\label{sec:kernel}

In \cite{HadPower2}, it was proved that the Hadamard codes obtained from Hadamard matrices $H(2,2^{t-1})$ over $\F_2$, with $t\geq 4$, have kernels of dimension $k\in \{1,2,\ldots, t-1,t+1\}$, and a construction of binary Hadamard codes of length $n=2^t$, with $t\geq 4$, for each one of these values was given. In \cite{HadAnyPower}, this result was generalized to Hadamard matrices $H(2,2^{t-1}s)$ over $\F_2$ showing that if there exists a binary Hadamard code of length $4s$, $s\not =1$ odd, then there exist binary Hadamard codes of length $n=2^ts$ for all $t\geq 2$, with kernel of dimension $k$ for all $k\in \{1,2,\ldots t-1\}$.
In this section, we include some results about the dimension of the kernel of GH codes with $q\not =2$.

\medskip
\begin{prop}\label{bounds}
Let $H(q,\lambda)$ be a $\gh$ matrix over $\F_q$, where $q=p^e$ and $p$ prime. Let $n=q\lambda=p^t s$ such that $\gcd(p,s)=1$.
Then $1 \leq \ker(C_H) \leq \ker_p(C_H) \leq 1+t/e$.
\end{prop}

\begin{IEEEproof}
Since $\K_p(F_H)$ is an $\F_p$-additive subcode of $F_H$, we have that $|\K_p(F_H)|=p^d$ for an integer $d$.
Moreover, by the properties of the kernel, since $F_H$ can be written as the union of cosets of $\K_p(F_H)$, we have that $p^d \mid p^t s$. Then, $d\leq t$ since $\gcd(p,s)=1$.
Therefore, $\ker_p(F_H)=d/e \leq t/e$, which proves that $0 \leq \ker(F_H) \leq \ker_p(F_H) \leq t/e$.
By Lemma~\ref{lemm:1}, the result follows.
\end{IEEEproof}

\begin{prop}\label{bounds-4}
Let $H(4,\lambda)$ be a $\gh$ matrix with $\lambda$ odd.
Then
$1 \leq \ker(C_H) \leq \ker_2(C_H) \leq 2$.
\end{prop}

\begin{IEEEproof}
Straigtforward from Proposition~\ref{bounds}
\end{IEEEproof}

Note that the upper bound in Proposition~\ref{bounds-4}  for $\ker(C_H)$ and $\ker_2(C_H)$ are tight as we can see by taking the unique normalized  $\gh$ matrix $H(4,1)=S_4$ given by the
multiplicative table of $\F_4$. In this case $\ker(C_H)=\ker_2(C_H)=2$. The lower bound is also tight since the unique normalized $\gh$ matrix  $H(4,3)$ has $\ker(C_H)=\ker_2(C_H)=1$.

\begin{lemm}\label{lemm:sq}
For $q>3$, there exists at least two versions of $S_q$ which are not translation equivalent. Moreover, they have a trivial intersection.
\end{lemm}

\begin{IEEEproof}
Recall that $S_q$ is the matrix given by the multiplicative table of $\F_q$ and let $S'_q$ be the matrix $S_q$ after a transposition of the second and third column. The entries of both matrices are elements from $\{0,1,\omega,\ldots,\omega^{q-2}\}$, where $\omega$ is a primitive element in $\F_q$. Both matrices $S_q$ and $S'_q$ are $\gh$ matrices, but they are not translation equivalent.
Indeed, $S_q\not=S_q'$ and if there exists a vector $\vv=(v_1,v_2,\ldots,v_q)$ such that $\vv+S_q=S'_q$, then $\vv \in S_q'$. Since $F_{S_q'}$ is a linear code over $\F_q$, $-\vv \in S_q'$. Let $\vw$ be
the row vector in $S_q$ that coincides with $-\vv$ in the last $q-3$ coordinates. Note that $\vw$ is completely determined by $\vv$, since $q>3$. Then, we have that $\vv+\vw =(0, v_2-v_3, v_3-v_2, 0, \ldots, 0)=\zero \in S_q'$, so $v_2=v_3$ and $\vv=\zero$.
\end{IEEEproof}

Let $\zero$, $\one$, $\bfomega^{(1)}$, $\ldots$, $\bfomega^{(q-2)}$ be the elements $0, 1,
\omega, \ldots, \omega^{q-2}$ repeated $\lambda$ times, respectively, where $\omega$ is a primite element in $\F_q$.

\begin{prop} \label{lem:switchq2}
For $q>2$, there exists a $\gh$ matrix $H(q,q)$ over $\F_q$ such that the $\gh$ code $C_H$ of length $n=q^2$ over $\F_q$ has $\ker(C_H)=2$.
\end{prop}

\begin{IEEEproof}
A GH matrix $H(q,q)$ over $\F_q$ such that $\ker(C_H)=2$ can be obtained by using
a switching construction. This kind of construction has already been used for Hadamard matrices over $\F_2$ in \cite{HadPower2}, \cite{HadAnyPower}.
Let $K$ be the linear subcode of $S^2=S_q \oplus S_q$ generated by the $q+1$ row vector $\vv$, that is,
$K=\langle\vv \rangle$, where $\vv=(\zero,\one,\bfomega^{(1)},\ldots,\bfomega^{(q-2)})$. Then, we take a coset $K+\vx \subset S^2$ such that $\vx \in S^2\backslash K$. Finally, we construct a matrix $H$ as $H=S^2 \backslash (K+\vx) \cup (K+\vx+\ve)$, where   $\ve=(0,\ldots,0,1,\ldots,1, 0,\ldots, 0)$ and the ones in $\ve$ cover the positions from $q+1$ to $2q$. Clearly, $F_H$ is nonlinear and $K\subseteq \K(F_H)$, so $K=\K(F_H)$ and $\ker(C_H)=2$ by Lemma~\ref{lemm:1}.

To prove that $H$ is a $\gh$ matrix, we just need to show that the multisets $\{ \vv_s-\vw_s : 1\leq s \leq q^2 \}$, for $\vw \in S^2\backslash (K+\vx)$ and $\vv \in K+\vx+\ve$, contains every element of $\F_q$ exactly $q$ times. We have that $\vw=\vk_1+\vx'$ and $\vv=\vk_2+\vx+\ve$ for some $\vk_1,\vk_2\in K$ and $\vx' \not \in K + \vx$, so $\vv-\vw=\vk''+\vx''+\ve$, where $\vk''\in K$ and $\vx''=\vx-\vx' \in S^2\backslash K$. It is clear that $\vk''+\vx'' \in S^2$ and each element of $\F_q$ appears once in each block of $q$ coordinates. Adding $\ve$ to $\vk''+\vx''$, we just change the order of the elements in the second block of $q$ coordinates, so $H$ fulfills the condition to be a $\gh$ matrix.
\end{IEEEproof}

\begin{ex} \label{ex:switch}
There are exactly two $\gh$ matrices $H(3,3)$ over $\F_3$, up to equivalence \cite{MT2000}.
One of them is the Sylvester $\gh$ matrix $S^2$ which has $\ker(C_{S^2})=3$.
The other one can be constructed using the switching construction given by the proof of Proposition \ref{lem:switchq2}. That is, first we consider the linear subcode $K=\langle  \vv \rangle \subset S^2$ over $\F_3$, where $\vv=(0,0,0,1,1,1,2,2,2)$. Then, we take for example the coset $K+\vx \not = K$, where $\vx=(0,1,2,0,1,2,0,1,2)$.
Finally, the matrix $H=S^2\backslash (K+\vx) \cup (K+\vx+\ve)$, where $\ve=(0,0,0,1,1,1,0,0,0)$, which is the matrix
$H$ given in Equation~(\ref{ex:H9}), is a $\gh$ matrix
such that the $\gh$ code $C_H$ of length $n=9$ has $\K(C_{H})=\langle  \one, \vv \rangle$, so $\ker(C_{H})=2$.
Note that the rows in boldface in Equation~(\ref{ex:H9}) indicate those that are in $K+\vx+\ve$.
\begin{equation} \label{ex:H9}
H=\left(
\begin{array}{ccccccccc}
0&0&0&0&0&0&0&0&0\\
\zero &\one & \two &\one&\two&\zero&\zero&\one&\two\\
0&2&1&0&2&1&0&2&1\\
0&0&0&1&1&1&2&2&2\\
\zero &\one&\two&\two&\zero&\one&\two&\zero&\one\\
0&2&1&1&0&2&2&1&0\\
0&0&0&2&2&2&1&1&1\\
\zero&\one&\two&\zero&\one&\two&\one &\two&\zero\\
0&2&1&2&1&0&1&0&2
\end{array}
\right)
\end{equation}
\end{ex}

\begin{theo} \label{theo:3.6}
For $q>2$, and any $h\geq 2$, except for $q=3$ and $h=2$, there exists a $\gh$ code $C_H$ of length $n=q^{h}$ over $\F_q$ with $\ker(C_H)=k$ if and only if $k\in \{1,2, \ldots, h+1 \}$. For $q=3$ and $h=2$, there exists a $\gh$ code $C_H$ of length $n=9$ over $\F_3$ with $\ker(C_H)=k$ if and only if $k\in \{2,3\}$.
\end{theo}

\begin{IEEEproof}
By Proposition \ref{bounds}, any GH code $C_H$ of length $n=q^{h}$ over $\F_q$ has $\ker(C_H) \in \{1,2,\ldots,t/e+1\}=\{1,2,\ldots,h+1\}$, since $h=t/e$.

For $q=3$ and $t=h=2$, there are exactly two nonequivalent $\gh$ matrices $H(3,3)$ over $\F_3$ \cite{MT2000}.
Both such matrices are also described in Example~\ref{ex:switch}, where it is shown that their corresponding $\gh$ codes have kernels of dimension 2 and 3, respectively. Therefore, there exists a GH code $C_H$ of length $n=9$ over $\F_3$ with $\ker(C_H)=k$ if and only if $k\in \{2,3\}$. Let $H_2$ and $H_3$ be the GH matrices $H(3,3)$ such that $\ker(C_{H_2})=2$ and $\ker(C_{H_3})=3$.
Note that $H_2$ can be constructed from $H_3$ using a switching construction, so $\K(C_{H_2}) \subseteq \K(C_{H_3})=C_{H_3}$.

For $q=3$ and $h=3$, using the Kronecker sum construction, we can obtain two GH matrices $H(3,3^2)$ as $S_q\oplus H_2$ and $S_q\oplus H_3$  with
 kernel of dimension 3 and 4, respectively, by  Corollary \ref{coro:2.2}. By Lemma \ref{lem:KroSumKernel}, the GH matrix $H(3,3^2)=S_q \oplus [H_2,H_3,\ldots,H_3]$ has $\ker(C_H)=2$. Again by Lemma \ref{lem:KroSumKernel}, the GH matrix $H(3,3^2)=S_q \oplus [\pi(H_2),H_3,\ldots,H_3]$ has $\ker(C_H)=1$, where $\pi$ is a permutation such that $\K(C_{H_2}) \cap \K(C_{\pi(H_2)}) = \langle \one \rangle$.

For $q>3$ and $h=2$, using the Kronecker sum construction, we can obtain GH matrices $H(q,q)=S^2=S_q \oplus S_q$ with $\ker(C_H)=3$ by Corollary \ref{coro:2.2}, and $H(q,q)=S_q \oplus [S'_q,S_q,\ldots, S_q]$, where $S'_q \cap S_q=\{\zero \}$, with $\ker(C_H)=1$ by Lemmas \ref{lem:KroSumKernel} and \ref{lemm:sq}. By Proposition \ref{lem:switchq2}, we can also obtain a GH matrix $H(q,q)$ such that $\ker(C_H)=2$ using a switching construction.

We have seen that the statement is true for any $q>3$ and $h=2$; and for $q=3$ and $h\in \{2,3\}$. Finally, by induction we will prove the result for any $h\geq 2$. Suppose it is true for
$n=q^{h-1}$, that is, there exists a GH code $C_{H_i}$ of length $q^{h-1}$ with kernel of dimension $i$ for all $i \in \{1,\ldots, h\}$. By Corollary \ref{coro:2.2}, the GH code obtained from $S_q \oplus H_i$ has a kernel of dimension $i+1$. Note that $H_1$ and $H_j$ for any $j$, $j\not =1$, have different ranks, which means that they are not translation equivalent. By Lemma \ref{lem:KroSumKernel}, the GH code corresponding to $S_q \oplus [H_1,H_j,\ldots, H_j]$ has a kernel of dimension 1.
\end{IEEEproof}

\begin{lemm} \label{lemm:minkernel}
Let $C_H$ be a $\gh$ code of length $n=q^h s$ over $\F_q$, where $s\not =1$ and $s$ is not a multiple of $q$. If $\ker(C_H)>1$, then $h \geq 2$.
\end{lemm}

\begin{IEEEproof}
If $\ker(C_H)>1$, then $\ker(F_H)\geq 1$ by Lemma \ref{lemm:1}. We can assume without loss of generality that
$\vv=(\zero, \one, \bfomega^{(1)}, \ldots, \bfomega^{(q-2)}) \in \K(F_H)$.
We consider all $n=q^h s$ coordinate positions divided into $q$ blocks of size $\lambda=q^{h-1}s$,
such that the coordinates of $\vv$ in each block coincide.

Let $\vx \in F_H\backslash \langle \vv \rangle$.
Let $\eta_\beta^{(i)}$ be the number of times an element $\beta \in \F_q$ appears in the coordinates of
the $i$th block of $\vx$. Since $\vx \in F_H$, we have that
\begin{equation} \label{eq:system1}
\eta_\alpha^{(1)}+\eta_\alpha^{(2)} +\cdots + \eta_\alpha^{(q)}=\lambda
\end{equation}  for all $\alpha \in \F_q$.
Adding these $q$ equations, we obtain
\begin{equation} \label{eq:sum}
\sum_{\alpha \in \F_q} \sum^q_{i=1} \eta_\alpha^{(i)}\equiv 0 \pmod{q}.
\end{equation}
Since $\vx +\omega^j \vv \in F_H$, we have that
\begin{equation} \label{eq:system2}
\eta_{\alpha}^{(1)}+\eta_{\alpha-\omega^j}^{(2)}+\eta_{\alpha-\omega^{j+1}}^{(3)}+\cdots +\eta_{\alpha-\omega^{j+q-2}}^{(q)}=\lambda
\end{equation}
for all $\alpha \in \F_q$ and $j \in \{0,\ldots,q-2\}$. For
a fixed $j\in \{0,\ldots,q-2\}$, adding the $q$ equations, we obtain $$\sum_{\alpha \in \F_q} (\eta_{\alpha}^{(1)}+\eta_{\alpha-\omega^j}^{(2)}+\eta_{\alpha-\omega^{j+1}}^{(3)}+\cdots +\eta_{\alpha-\omega^{j+q-2}}^{(q)})=\sum_{\alpha \in \F_q} \sum^q_{i=1} \eta_{\alpha}^{(i)} \equiv 0 \pmod{q}.$$
The size of each block is $\lambda$, so we also have that
\begin{equation} \label{eq:system3}
\sum_{\alpha \in \F_q} \eta^{(i)}_\alpha=\lambda
\end{equation} for all $i\in \{1,\ldots,q\}$,
and adding these $q$ equations, we obtain the same relation as in Equation~(\ref{eq:sum}).
From these $q(q+1)$ equations, given by Equations~(\ref{eq:system1}), (\ref{eq:system2}), and (\ref{eq:system3}), we have seen that there are $q$ which are linear dependent, and the rest are
 linear independent. The fact that they are linear independent is easily seen by writing the equations in matrix form. The system of equations has a unique solution $\eta_\alpha^{(i)}=\lambda/q$ for all $\alpha\in \F_q$ and $i \in \{1,\ldots,q\}$,
 which means that $\lambda$ has to be a multiple of $q$, so $h\geq 2$.
\end{IEEEproof}

\begin{lemm} \label{lemm:maxkernel}
Let $C_H$ be a $\gh$ code of length $n=q^h s$ over $\F_p$, where $s\not =1$ and $s$ is not a multiple of $q$. Then $\ker(C_H) \leq h$.
\end{lemm}

\begin{IEEEproof}
If $C_H$ has length $n=qs$, that is, if $h=1$, then $\ker(C_H)=1$ and the statement is true by Lemma \ref{lemm:minkernel}.
If $h>1$, assume that $C_H$ of length $n=q^hs$ has $\ker(C_H)=k>1$.
Let $\vv \in \K(C_H)\backslash \langle \one \rangle$. Let $I$ be the set of coordinates where the vector $\vv$ is equal to one.
We puncture $C_H$ eliminating all coordinates outside $I$.
There are exactly $q$ copies of each vector, since for any $\vx \in C_H$, the vector $\vx+\bfomega^{(j)}-\bfomega^{(j)} \vv$ is the same
as $\vx$ in the coordinates of $I$, for all $j\in \{0,\ldots,q-2\}$. Note that $\vx+\bfomega^{(j)}-\bfomega^{(j)} \vv \in C_H$, since
$\bfomega^{(j)}-\bfomega^{(j)} \vv \in \K(C_H)$. Let $C$ be the new code without repeated vectors. Note that $C$ has length $n=q^{h-1}s$
and a kernel of dimension $k-1$, since the independent vectors in $\K(C_H)$ are still independent in
$\K(C)$ except $\vv$ which coincides with $\one$. Using the same argument as in the proof of Lemma \ref{lemm:minkernel},
it is easy to see that $C$ is a $\gh$ code. By using the induction hypothesis $k-1\leq h-1$, so $k\leq h$.
\end{IEEEproof}

\begin{theo}
Let $q=p$, $p$ prime, $s \not =1$ such that $s$ is not a multiple of $q$, and $h\geq 2$. If there exists a $\gh$ matrix $H(q,s)$, then there exist $\gh$ codes $C_H$ of length $n=q^h s$ over $\F_q$ with $\ker(C_H)=k$ if and only if $k\in \{1,2, \ldots, h\}$.
\end{theo}

\begin{IEEEproof}
If there exists a $\gh$ matrix $H(q,s)$, then the corresponding $\gh$ code $C_H$ of length $n=qs$ has $\ker(C_H)=1$, by Lemma \ref{lemm:minkernel} and Proposition \ref{bounds}. The upper bound for the dimension of the kernel is given by Lemma \ref{lemm:maxkernel}.
Following the same argument as in the last part of the proof of Theorem~\ref{theo:3.6}, we have the statement.
\end{IEEEproof}

\section{Rank of $\gh$ codes}
\label{sec:rank}

In \cite{HadPower2}, it was proved that the Hadamard codes obtained from Hadamard matrices $H(2,2^{t-1})$ over $\F_2$, with $t\geq 3$, have ranks $r\in \{t+1,\ldots, n/2\}$, and a construction of binary Hadamard codes of length $n=2^t$ for each one of these values was given. In \cite{HadAnyPower}, it was shown that if there exists a binary Hadamard code of length $4s$, $s\not =1$ odd, which
always has rank $n-1$ \cite{AssKey}, then there exist binary Hadamard codes of length $n=2^ts$ for all $t\geq 3$, with rank $r$ for
all $r\in \{4s+t-3,\ldots, n/2\}$.

In this section, we give an upper bound for the rank proving that any $\gh$ matrix $H(q,\lambda)$ over $\F_q$ with $q>2$ is self-orthogonal, except $q=3$ and $\gcd(\lambda,3)=1$. Moreover, for some particular cases, we specify lower and upper bounds on the rank, once the dimension of the kernel is given. Finally, $\gh$ codes having all different ranks between these bounds are constructed for some of these cases.

\medskip
For vectors over $\F_q$, $q=p^e$ and $p$ prime, we have the {\it Euclidean inner product}.  Namely
$[\vv,\vw] = \sum^n_{i=1} v_i w_i$ for any $\vv, \vw \in \F_q^n$.   If $C$ is a code over $\F_q$ of length $n$, then we define the {\it Euclidean orthogonal code} as $C^\perp = \{ \vv \ :\  [\vv,\vw] = 0$ for all $\vw \in C\}$. Note that $C^\perp$ is always linear over $\F_q$ whether $C$ is or not. In fact, if $C$ is a nonlinear code, then $C^\perp=\langle C \rangle^\perp$. Moreover, we say that $C$ is {\it Euclidean self-orthogonal} if $C\subseteq C^\perp$ and {\it Euclidean self-dual}
if $C=C^\perp$.

\begin{lemm} \label{sum}
Let $H(q,\lambda)$ be a $\gh$ matrix over $\F_q$.  Then ${\bf 1} \in  {\cal R}(F_H)^\perp$.
\end{lemm}
\begin{IEEEproof}
The sum of the elements of $\F_q$, $q=p^e$ and $p$ prime, is 0. Indeed, the elements of the finite field $\F_q$ are the roots of the polynomial $x^q-x$ and the sum of all these roots is the coefficient of $x^{q-1}$, which is zero (except for $q=2$).
Then, since each row vector $\vv$ in $H$ has all the elements in $\F_q$ repeated $\lambda$ times, we have, for $q\not=2$, that $[{\bf 1}, \vv ] = \lambda 0 =0$ in $\F_q$ for all rows $\vv $ in $H$. When $q=2$, since $\lambda$ is always even, we also obtain $[{\bf 1}, \vv ] =0$ in $\F_2$.
\end{IEEEproof}

\begin{lemm} \label{lemm:onezeros}
Let $H(q,\lambda)$ be a normalized $\gh$ matrix over $\F_q$.
Then $(1,0,\dots,0) \in  {\cal R}(F_H)^\perp$.
\end{lemm}
\begin{IEEEproof}
If the matrix is normalized, then the first coordinate is 0 and the result follows.
\end{IEEEproof}

\begin{prop} \label{prop:boundRank}
Let $H(q,\lambda)$ be a normalized $\gh$ matrix over $\F_q$.
Then $\rank(C_H) \leq n-1$, where $n=q\lambda$.
\end{prop}
\begin{IEEEproof}
By the previous two lemmas, we have that the dimension of ${\cal R}(F_H)^\perp=F_H^\perp$ is at least 2. Then, $\rank(F_H) =n- \rank(F_H^\perp)\leq n-2$.
By Lemma~\ref{lemm:1}, $\rank(C_H) \leq n-1$.
\end{IEEEproof}

The upper bound given by Proposition \ref{prop:boundRank} can be met, for instance, for the $\gh$ matrices given in Examples \ref{ex:NoSelfOZ2} and \ref{ex:NoSelfOZ3}.

\begin{ex}  \label{ex:NoSelfOZ2}
The normalized Hadamard matrices $H(2,2\mu)$ over $\F_2$ satisfy that $\rank(F_H)=4\mu-2$, so  $\rank(C_H)=4\mu-1$, as long as $\mu$ is odd \cite{AssKey}.
\end{ex}

\begin{ex} \label{ex:NoSelfOZ3}
The unique normalized $\gh$ matrix $S_3=H(3,1)$ over $\F_3$, given by the multiplicative table of $\F_3$, has $\rank(C_H)=2$.
Consider the normalized $\gh$ matrix $H(3,2)$ over $\F_3$, given by (\ref{ex:upperBound}). It is easy to verify that there is an unique normalized $\gh$ matrix $H(3,2)$ over $\F_3$, up to equivalence. In this case, $\rank(F_H )=4$, so $\rank(C_H)=5$.
\begin{equation} \label{ex:upperBound}
H(3,2)=\left(
\begin{array}{cccccc}
0&0&0&0&0&0\\
0&0&1&1&2&2\\
0&1&2&0&1&2\\
0&1&0&2&2&1\\
0&2&1&2&1&0\\
0&2&2&1&0&1
\end{array}
\right)
\end{equation}
There is also an unique normalized $\gh$ matrix $H(3,4)$ over $\F_3$, up to equivalence \cite{deLauney}, which has $\rank(C_H)=11$.
\end{ex}


Note that the $\gh$ matrices from the previous two examples generate codes that are not self-orthogonal. However, there are many others $\gh$ matrices such that they do lead to self-orthogonal codes. For example, it is known that the Hadamard matrices $H(2,2\mu)$ over $\F_2$ generate self-orthogonal codes when $\mu$ is even \cite{AssKey}.
Moreover, in \cite{Harada}, it is shown by computer that the code generated by any $\gh$ matrix $H(4,4)$ over $\F_4$ is  self-orthogonal.

\medskip
What we do next is to show that any $\gh$ matrix $H(q,\lambda)$ over $\F_{q}$, with $q>3$, or $q=3$ and $\gcd(3,\lambda) \not =1$, generates a self-orthogonal code.

Let $\Z_{2p}$ be the ring of integers modulo $2p$. A monic polynomial $f(x)\in \Z_{2p}[x]$ is called monic basic irreducible (primitive) if computing it modulo 2 we obtain  an irreducible (primitive) polinomial ${\bar f}(x)\in \F_2[x]$. We define the ring $R_{(2p)^e} = \Z_{2p}[x]/\langle f(x)\rangle$, where $f(x)$ is a monic basic irreducible polynomial of degree $e$ (which always exists). The ring $R_{(2p)^e}$ has a maximal ideal $\langle p\rangle$ and its residue field $R_{(2p)^e}/\langle p\rangle$ is isomorphic to the field $\F_{p^e}$  of order $p^e$. In the ring $R_{(2p)^e}$ there exists a nonzero element $\omega$ of order $p^e-1$, which is a root of a basic primitive polynomial $f(x)$ of degree $e$ over $\Z_{2p}$ and $R_{(2p)^e}=\Z_{2p}[\omega]$.  Let $T =\{0,1,\omega,\ldots,\omega^{p^e-2}\}$, then any element of $R_{(2p)^e}$ can be written uniquely, using a $p$-adic representation, as $a+pb$, where $a,b\in T$. Analogously, the elements in $\F_{p^e}$ are those in ${\bar T}= \{0,1,{\bar \omega},\ldots,{\bar \omega}^{2^e-2}\}$, where ${\bar \omega}$ is a root of the primitive polynomial ${\bar f}(x)\in \F_p[x]$.

\begin{lemm}\label{first}
If $p=2$ and $e > 1$, then $\sum_{\alpha \in T} \alpha^2 \equiv 0 \pmod{2}$.
\end{lemm}
\begin{IEEEproof}
Since $\gcd(2,2^e-1)=1$, every element of the multiplicative cyclic group $\langle \omega\rangle$ can be written as a square element of $\langle \omega\rangle$. Hence,  $\sum_{\alpha \in T} \alpha^2 =\sum_{\alpha \in T} \alpha$. Taking modulo 2 and noting that the sum of all elements in ${\bar T}$ is zero (indeed, the elements in ${\bar T}$ are the roots of $x^{2^e-1}-x\in Z_2[x]$ and the addition of all of them is the coefficient of $x^{2^e-2}$, which is zero for $e\geq 2$) we obtain, using the 2-adic presentation, $\sum_{\alpha \in T} \alpha = 2b$ with $b\in T$. Now, we show that $2b=0$. Squaring each side the last equation $\sum_{\alpha \in T} \alpha = 2b$ we obtain $\sum_{\alpha \in T} \alpha^2 + \sum_{i\not=j; i,j \in \{0,\ldots,2^e-2\}} \omega^i\omega^j = 0$. Hence, $2b + \sum_{k\in \{0,\ldots,2^e-2\}}\lambda_k\omega^k =0$, where $\lambda_k$ stands for the number of ways we can split $k$ or $2^e-1+k$ in two different addends $ i,j \in \{0,\ldots,2^e-2\}$. It is easy to see that $\lambda_k= 2^e-2$ for all $k\in \{0,\ldots,2^e-2\}$  and so $2b + (2^e-2)\sum_{\alpha \in T} \alpha =0$. Finally, $(2^e-1)2b=0$. Therefore $2b=0$, which proves the statement.
\end{IEEEproof}

\begin{lemm} \label{firstA}
Let $p$ be an odd prime.  If $p^e>3$, then $\sum_{x\in \F_{p^e}} x^2 \equiv 0 \pmod{2p}$. If $p^e=3$, then $\sum_{x\in \F_{3}} x^2 \equiv 2 \pmod{3}$.
\end{lemm}

\begin{IEEEproof}
For $p^e=3$ the result is straightforward. For $p^e>3$, let $a \in \F_{p^e}$ such that $a^2 \neq 0,1$. As long as $p^e>3$ this is easily done.  Then $$\sum_{x\in \F_{p^e}} x^2 = \sum_{x\in \F_{p^e}} (ax)^2
= a^2 \sum_{x\in \F_{p^e}} x^2.$$  This gives that $\sum_{x\in \F_{p^e}} x^2  =0$ in $\F_{p^e}.$

Now split the nonzero elements of $\F_{p^e}$ into two disjoint sets $A$ and $B$ such that $x \in A$ if and only if $-x \in B.$ Since $p\neq 2$, the elements $a$ and $-a$ are always distinct.  Then $\sum_{x\in A} x^2  =\sum_{x\in B} x^2   $ and $ \sum_{x\in \F_{p^e}} x^2 = \sum_{x\in A} x^2  +\sum_{x\in B} x^2.$
This gives that $2 \sum_{x\in A} x^2 = 0$ and so $\sum_{x\in A} x^2  = 0$ in $\F_{p^e}.$  Then we have that
   $$ \sum_{x\in \F_{p^e}} x^2 = 2  \sum_{x \in A } x^2 \equiv 0 \pmod{2p}.$$
\end{IEEEproof}

Let $\vv,\vw \in \F_{p^e}^n$. The Euclidian inner product $[\vv,\vw] = \sum v_iw_i$ is computed in $\F_{p^e}$, but we can substitute the elements $\{0,1,{\bar \omega},\ldots,{\bar \omega}^{p^e-2}\}$ of $\F_{p^e}$ by the corresponding representatives $\{0,1,\omega,\ldots,\omega^{p^e-2}\}$ in $R_{(2p)^e}$ and we call $[\vv,\vw]_{2p} \in R_{(2p)^e}$ the result we obtain on computing the above inner product.

\begin{lemm}\label{third}
Let $H(p^e,\lambda)$ be a $\gh$ matrix over $\F_{p^e}$. Let $\vv$ and $\vw$ be two rows of $H$. If $p^e>3$, then
$[\vv,\vv]_{2p}=0$, $[\vw,\vw]_{2p}=0$ and $[\vv+\vw,\vv+\vw]_{2p}=0$. If $p^e=3$, then $[\vv,\vv]=[\vw,\vw]=[\vv+\vw,\vv+\vw]=2\lambda$.
\end{lemm}
\begin{IEEEproof}
We have that every element of $\F_{p^e}$ appears $\lambda$ times in $\vv$, $\vw$ and $\vv+\vw.$
Then the result follows from Lemmas~\ref{first} and \ref{firstA}.
\end{IEEEproof}

We can now prove our desired result.

\begin{theo}\label{big}
Let $H(p^e,\lambda)$ be a $\gh$ matrix over $\F_{p^e}$, with $p^e>3$, or $p^e=3$ and $\lambda$ a multiple of 3.
Then ${\cal R}(F_H)$ is self-orthogonal.
\end{theo}
\begin{IEEEproof}
Let $\vv,\vw$ be rows of $H$. For $p^e>3$ we know that $[\vv,\vv]_{2p} = [\vv+\vw,\vv+\vw]_{2p}=0$ by Lemma~\ref{third}.
Then, we have
\begin{eqnarray*}
0 = [{\vv}+{\vw},{\vv}+{\vw}]_{2p} &=& \sum ({v_i} + {w_i})^2 \\
&=& \sum {v_i}^2 + {w_i}^2 + 2 \sum {v_i} {w_i} \\
&=& [\vv,\vv]_{2p}+ [\vw,\vw]_{2p} + 2 \sum {v_i} {w_i} \\
&=& 2 \sum {v_i} {w_i}.
\end{eqnarray*}
Since $0 = 2 \sum {v_i} {w_i}$ in $R_{(2p)^e}$, we have that $\sum v_i w_i \equiv 0 \pmod{p}.$
Hence $[\vv,\vw]=0$, which gives that ${\cal R}(F_H)$ is a self-orthogonal code.

For $p^e=3$ and $\lambda$ a multiple of 3 we have
$$
2\lambda = [{\vv}+{\vw},{\vv}+{\vw}]=[\vv,\vv]+[\vw,\vw]+2 \sum {v_i} {w_i}=2\lambda+2\lambda+2 \sum {v_i} {w_i},
$$
so $2\lambda=2 \sum {v_i} {w_i} \pmod{3}$. If $\lambda$ is a multiple of 3, we have $\sum {v_i} {w_i} \equiv 0 \pmod{3}$ and $[\vv,\vw]=0$, which gives that ${\cal R}(F_H)$ is a self-orthogonal code.
\end{IEEEproof}

This leads to the following corollary.

\begin{coro} \label{coro:upperBoundRank}
Let $H(q,\lambda)$ be a normalized $\gh$ matrix  over $\F_{q}$, with $q>3$, or $q=3$ and $\lambda$ a multiple of 3. Then $\rank(C_H) \leq \lfloor n/2 \rfloor$, where $n=q\lambda$.
\end{coro}
\begin{IEEEproof}
By Theorem~\ref{big}, we have that ${\cal R}(F_H)$ is self-orthogonal.  Moreover, by Lemmas \ref{sum} and \ref{lemm:onezeros}, we have that $(1,0,\dots,0)$ and ${\bf 1}$ are in ${\cal R}(F_H)^\perp=F_H^\perp$ but not in ${\cal R}(F_H).$ Then
$\rank(F_H^\perp) \geq 2+ \rank(F_H)$ and $n -  \rank(F_H)\geq 2 + \rank(F_H)$ which gives $\rank(F_H) \leq n/2-1.$ By Lemma~\ref{lemm:1}, $\rank(C_H) \leq n/2$.\end{IEEEproof}

\begin{coro} \label{coro:allranks}
Let $H(q,q^{h-1})$ be a $\gh$ matrix over $\F_q$, with $q>3$ and $h \geq 1$, or
$q=3$ and $h\geq 2$. Then $\rank(C_H) \in \{ h+1, \ldots, \lfloor q^h/2 \rfloor \}$.
\end{coro}

\begin{IEEEproof}
For $q=3$ and $h\geq 2$, and for $q>3$ and $h\geq 1$,
by Corollary \ref{coro:upperBoundRank}, we have that $\rank(C_H)\leq \lfloor n/2 \rfloor= \lfloor q^h/2 \rfloor$. If $H=S^h$, where $S^h$ is the generalized Sylvester Hadamard matrix of order $q^h$, then $C_H$ is a linear code of length $n=q^h$ over $\F_q$, so $\rank(C_H)=h+1$ and the result follows.
\end{IEEEproof}

\begin{prop} \label{prop:boundsGivenKernel}
Let $H(q,q^{h-1})$ be a $\gh$ matrix over $\F_q$, with $q>2$ and $h \geq 1$.
Let $r=\rank(C_H)$ and $k=\ker(C_H)$. Then,
\begin{description}
\item{(i)} if $k=1$, then $h+2 \leq r \leq \lfloor q^h/2 \rfloor$;
\item{(ii)} if $2\leq k \leq h$, then $h+2 \leq r \leq k+q^{h+1-k}-1$;
\item{(iii)} if $k=h+1$, then $r=h+1$.
\end{description}
\end{prop}

\begin{IEEEproof}
First note that when $q=3$ and $h=1$, the $\gh$ matrix $H(3,1)$ is unique up to equivalence, and generates a linear code $C_H$, so $r=k=h+1=2$.

For all other cases, we can use the following argument.
We know that $\K(C_H)$ is the largest linear subspace in $C_H$ such that
$C_H$ can be written as the union of cosets of $\K(C_H)$. Since there are $q^{h+1-k}$ cosets in $C_H$,
and $r$ is maximum when each coset contributes an independent vector, we have that $r\leq k+q^{h+1-k}-1$. This same argument was used in \cite{HadPower2},  \cite{HadAnyPower} for binary Hadamard codes and in \cite{merce3} for 1-perfect codes. If $k=1$, we have that $r\leq \lfloor q^h/2\rfloor$, by Corollary \ref{coro:allranks}, and because $q^h/2 < k+q^{h+1-k}-1=q^h$. Finally, if $k=h+1$, then $C_H$ is linear, so $r=h+1$. For the lower bound, just note that if $k \leq h$, then $C_H$ is nonlinear, so
$r\geq h+2$.
\end{IEEEproof}

The bounds given by Corollary \ref{coro:allranks} and Proposition \ref{prop:boundsGivenKernel} can be met, for example, for the $\gh$ matrices $H(4,4)$ over $\F_4$, as we can see in Example \ref{ex:RanksN16}.

\begin{ex} \label{ex:RanksN16} It is known that there are exactly 226 nonequivalent $\gh$ matrices $H(4,4)$ over $\F_4$ having $\rank(C_H) \in \{3,4,5,6,7,8\}$ \cite{Harada}. Moreover, the number of such $\gh$ matrices with respect to the pair ($\rank(C_H)$, $\ker(C_H)$) is given by Table \ref{table:rkN16}. From this table, it is easy to see that the bounds on $\rank(C_H)$, once $\ker(C_H)$ is given, satisfy Proposition \ref{prop:boundsGivenKernel}.
\begin{table}[ht] \caption{Parameters $\rank(C_H)$ and $\ker(C_H)$ of $\gh$ matrices $H(4,4)$ over $\F_4$.}
\begin{center} \begin{tabular}{r | c c c c c c}
   & \multicolumn{6}{c}{$\rank(C_H)$} \\
$\ker(C_H)$  & 3 & 4 & 5 & 6 & 7 & 8 \\
\hline
3 & 1 &   &   &     \\
2 &   & 7  & 8 &     \\
1 &   & 3  & 92  & 55 & 57 & 3    \\
\end{tabular}   \label{table:rkN16}
\end{center}
\end{table}
\end{ex}

\begin{coro}
Let $H(q,q^{h-1})$ be a $\gh$ matrix over $\F_q$ with $q>2$ and $h\geq 1$. If
$C_H$ is self-dual, then $q$ is even and $\ker(C_H)=1$.
\end{coro}

\begin{IEEEproof}
If $C_H$ is self-dual, then $\rank(C_H)=n/2$.
Since $n/2$ must be an integer, $q$ must be even.
Therefore, the result follows from Proposition \ref{prop:boundsGivenKernel}.
\end{IEEEproof}

\begin{prop} For $q>2$, and any $h\geq 2$ such that $\lceil (h+2)/2 \rceil \leq k \leq h$, there exists a $\gh$ code $C_H$ of length $n=q^h$ over $\F_q$ with $\ker(C_H)=k$ and $\rank(C_H)=r$ if and only if $r\in \{h+2, \ldots, k+q^{h+1-k}-1 \}$.
\end{prop}

\begin{IEEEproof}
We will see that the corresponding $\gh$ matrices $H(q,q^{h-1})$ can be generated by using a switching construction.
Let $S^h$ be the Sylvester $\gh$ matrix $H(q,q^{h-1})$.
We can assume without loss of generality that $S^h$ is generated by the vectors $\vv_1, \ldots, \vv_{h}$ of length $n=q^h$,
where 
$$\vv_i=(\zero_{i}, \one_{i}, \bfomega_{i}^{(1)}, \ldots, \bfomega_{i}^{(q-2)}, \ldots, \zero_{i}, \one_{i}, \bfomega_{i}^{(1)}, \ldots, \bfomega_{i}^{(q-2)} ),$$
$\zero_{i}$, $\one_{i}$, $\bfomega_{i}^{(1)}$, $\ldots$, $\bfomega_{i}^{(q-2)}$ are the elements $0, 1,
\omega, \ldots, \omega^{q-2}$ repeated $q^{h-i}$ times, respectively, and $\omega$ is a primite element in $\F_q$, for all $i\in \{1,\ldots,h\}$.
Let $K$ be the linear subcode of $S^h$ generated by the $k-1$ row vectors $\vv_1, \ldots, \vv_{k-1}$.
Note that all $n=q^h$ coordinates are naturally divided into $q^{k-1}$ groups of
size $q^{h-k+1}$, which will be referred to as blocks, such that the
columns of $K$ in a block coincide.

The rows of $S^h$ can be partitioned into $q^{h-k+1}$ cosets of $K$.
We can take any coset $K+\vx_1 \subset S^h$ such that $\vx_1\in S^h\backslash K$ and construct a matrix $H$ as
$H=S^h \backslash (K+\vx_1) \cup (K+\vx_1+\ve_1)$, where $\ve_1$ is the vector of length $n=q^h$ with ones in the positions given by the second block and zeros elsewhere. Clearly, $\rank(F_H)=h+1$ and $K\subseteq \K(F_H)$. Moreover, it is easy to prove that $K=\K(F_H)$. Therefore, $\ker(F_H)=k-1$, which gives $\ker(C_H)=k$ by Lemma \ref{lemm:1}.

This process can be performed several times. Let $\ve_s$ be the vector of length $n=q^h$ with ones in the positions given by the $(s+1)$th block and zeros elsewhere, for all $s \in \{1,\ldots, q^{h+1-k}+k-h-2\}$.
Since $\lceil (h+2)/2 \rceil \leq k$, we have that $k-1 \geq h-k+1$. Therefore,
$q^{k-1} \geq q^{h-k+1} \geq q^{h+1-k}+k-h-2$, so there are enough blocks to define vector $e_s$.
Let $\vx_1, \vx_2,\ldots, \vx_s \in S^h \backslash K$ such that $\bigcap_{j=1}^s (K+\vx_j)=\emptyset$.
Then, we can construct the matrix $$H=(S^h \backslash \bigcup_{j=1}^s (K+\vx_j)) \cup \bigcup_{j=1}^s(K+\vx_j+\ve_j).$$
Again, it is clear that $\rank(C_H)=h+1+s$ and $\ker(C_H)=k$.
To prove that $H$ is a $\gh$ matrix, we can use the same argument as in the proof of Proposition \ref{lem:switchq2}.
\end{IEEEproof}

\section{Conclusions}
\label{sec:conclusion}

We have established lower and upper bounds for the dimension of the kernel and rank of codes
constructed from $\gh$ matrices over $\F_q$. For some cases, we proved that these bounds are tight, by constructing
$\gh$ matrices for each possible rank, once the dimension of the kernel is given.
Further research on this topic would include
giving the construction of a $\gh$ matrix $H(q,\lambda)$ over $\F_q$ for each possible pair $(\rank(C_H),\ker(C_H))$
or providing similar results for the parameters $p$-rank and $p$-kernel of these codes.
Another direction of future research could be to focus on the rank and kernel of $\F_p$-additive $\gh$ codes, which are the ones
having the $p$-rank equal to the $p$-kernel, or in a more general way, study $K$-additive $\gh$ codes, where $K$ is a subfield of $\F_q$.

\nocite{}

\end{document}